\documentclass[sigconf]{acmart}

\usepackage{tabularray}

\AtBeginDocument{%
  \providecommand\BibTeX{{%
    \normalfont B\kern-0.5em{\scshape i\kern-0.25em b}\kern-0.8em\TeX}}}

\copyrightyear{2022}
\acmYear{2022}
\setcopyright{rightsretained}
\acmConference[KDD '22]{Proceedings of the 28th ACM SIGKDD Conference on Knowledge Discovery and Data Mining}{August 14--18, 2022}{Washington, DC, USA}
\acmBooktitle{Proceedings of the 28th ACM SIGKDD Conference on Knowledge Discovery and Data Mining (KDD '22), August 14--18, 2022, Washington, DC, USA}
\acmDOI{10.1145/3534678.3539054}
\acmISBN{978-1-4503-9385-0/22/08}

\begin{document}

\title{Affective Signals in a Social Media Recommender System}

\author{Jane Dwivedi-Yu}

\affiliation{%
  \institution{Meta AI}
  \city{London}
  \country{UK}
}
  \email{janeyu@fb.com}

\author{Yi-Chia Wang}
\affiliation{%
  \institution{Meta AI}
  \city{Menlo Park}
  \state{CA}
  \country{USA}
  }
\email{yichiaw@fb.com}

\author{Lijing Qin}
\affiliation{%
  \institution{Meta AI}
   \city{Seattle}
  \state{WA}
  \country{USA}
}
\email{qinlijing@fb.com}

\author{Cristian Canton-Ferrer}
\affiliation{%
  \institution{Meta AI}
  \city{Seattle}
  \state{WA}
  \country{USA}
  }
  \email{ccanton@fb.com}

\author{Alon Y. Halevy}
\affiliation{%
  \institution{Meta AI}
  \city{Menlo Park}
  \state{CA}
  \country{USA}
  }
\email{ayh@fb.com}

\renewcommand{\shortauthors}{Jane Dwivedi-Yu et al.}

\begin{abstract}
  
  People come to social media to satisfy a variety of needs, such as being informed, entertained and inspired, or connected to their friends and community. Hence, to design a ranking function that gives useful and personalized post recommendations, it would be helpful to be able to predict the affective response a user may have to a post (e.g., entertained, informed, angered). This paper describes the challenges and solutions we developed to apply Affective Computing to social media recommendation systems.
     
  We address several types of challenges. First, we devise a taxonomy of affects that was small (for practical purposes) yet covers the important nuances needed for the application. Second, to collect training data for our models, we balance between signals that are already available to us (namely, different types of user engagement) and data we collected through a carefully crafted human annotation effort on 800k posts. We demonstrate that affective response information learned from this dataset improves a module in the recommendation system by more than 8\%. Online experimentation also demonstrates statistically significant decreases in surfaced violating content and increases in surfaced content that users find valuable.
  

\end{abstract}


\maketitle

\section{Introduction}

Social media platforms have become a common means of interacting and connecting with others as well as finding interesting, informing, and entertaining content \cite{craig2021can, uhls2017benefits, khan2014social}. Users of those platforms depend on the ranking systems of the recommendation systems to show them information they will be most interested in, provide them with positive experiences, and safeguard them against offensive material. Users may also look for online content that will help them change or enhance their current affective state \cite{myrick2015emotion, van2016online}, and social media is indeed rich in these signals \cite{kivran2011network, kramer2014experimental, thelwall2010data}. Conceivably, Affective Computing, a field that develops methods to predict user's affects in a particular application context (e.g., customer support interaction, online learning), can contribute to the suite of methods used by recommender systems to ensure that users are having the best experiences possible \cite{tkalcic2012affective, zheng2013role,DBLP:journals/corr/abs-1903-01728,mishra20,DBLP:conf/acii/FP19}. This paper describes the challenges and solutions we developed to apply Affective Computing in the context of Facebook's ranking algorithm.  

The first contribution of this paper is translating the vision of applying Affective Computing into a well-specified technical problem. Doing so involved several challenges. The first challenge was to understand where in the complex recommendation system an affective prediction can be useful for ranking and how to combine it with other ranking signals. The second challenge was to devise a set of criteria for determining which affects pertain to recommender systems (e.g., inspiration, entertainment, sadness, fear) and to apply these criteria to decide on a reasonably short list of affects to operationalize.  

The second set of contributions concern the operationalization of Affective Computing where we developed models for predicting the potential affective response a user may have to a post. We began by developing methods for training affect classifiers based on users' engagement (reactions, shares, outbound clicks, negative user feedback, and comments) on the platform, which carries an important signal about their affective response towards content. However, engagement alone does not suffice, because a single response can sometimes indicate different intents given different contexts. For instance, consider the sad or sorry reaction used in response to a post. In some cases, the reaction may simply indicate the user's emotional support, but in other cases, it may indicate the user's sadness and a preference to never see that type of content again. Hence, we also supplemented the signal based on engagement with a carefully constructed human annotation pipeline to directly label affective response. Using this data, we developed a model for predicting affective signals that was then injected into the ranking algorithm of the recommendation system, leading to substantial improvements of several metrics.   

The paper is organized as follows. Section~\ref{section:definition} describes where Affective Computing  can contribute to a recommendation system and how we decided which affects to operationalize.  Section~\ref{section:methods} describes how we create training data for our classifiers and the modeling architecture and features for predicting affective signals. Section~\ref{section:evaluation} provides an analysis of both the dataset and model when incorporated into the recommendation system, describing experiments that validate some of the choices we made in this work. Section~\ref{section:related} describes related work and Section~\ref{section:conclusions} concludes and points to future avenues of research.


\section{Problem definition}
\label{section:definition}

This section begins by reviewing the basic aspects of recommendation systems that are relevant to our discussion (Section~\ref{section:recsystems}). We then define the concept of affective response and discuss how it can be incorporated into the recommendation system (Section~\ref{section:af}). Finally, we discuss how we chose the affects to model in our work (Section~\ref{section:taxonomy}). 


\subsection{Recommendation systems}
\label{section:recsystems}
\begin{figure}[tp]
  \centering
  \includegraphics[width=\linewidth]{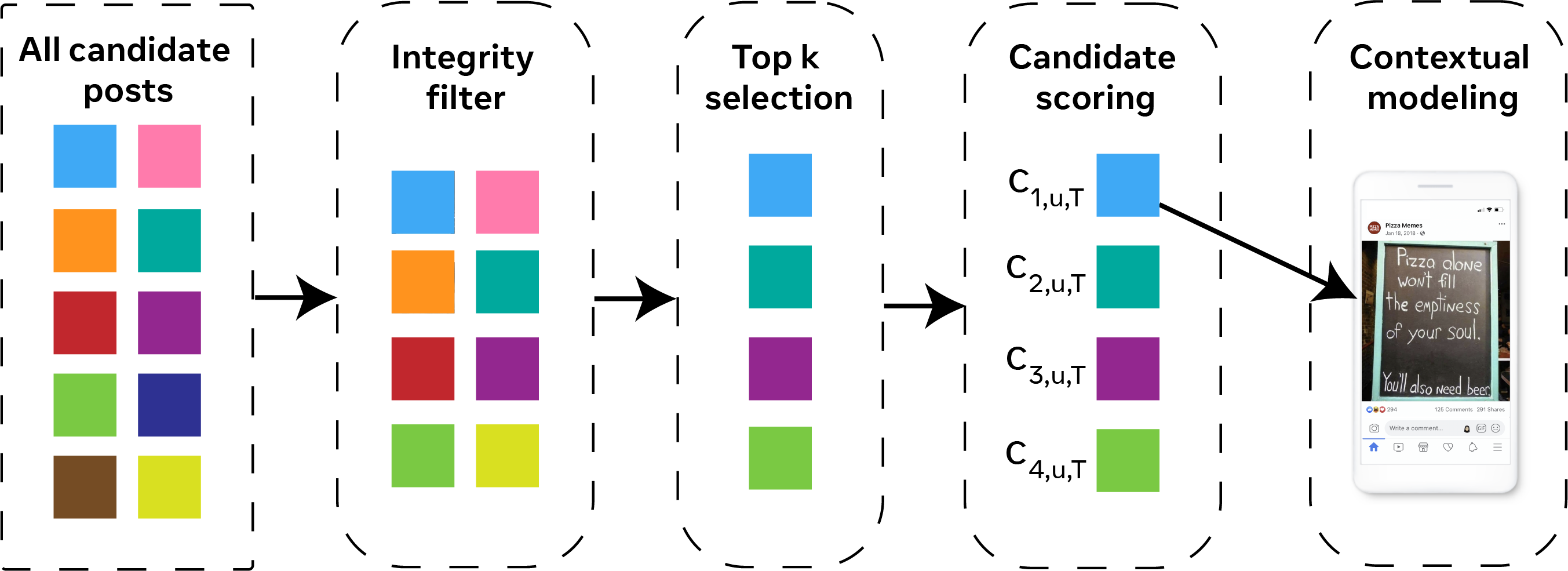}
  \caption{Example of a recommender system. The first step removes any violating content from the pool of candidate items. The top 500 or so of the remaining posts get scored on whether post p would produce value for the user u at time T, denoted  $c_{p,u,T}$.  Lastly, the selected posts undergo a re-ranking step to ensure the feed contains enough diversity.}
  \label{fig:recommender}
\end{figure}

 The recommendation system is the component responsible for deciding which posts will be shown to a user and in what order. A typical architecture of a recommendation system inspired by ~\cite{facebook-engineering-blog} is shown in Figure~\ref{fig:recommender}. 

The recommendation system starts with a pool of available posts. These posts may have been posted by the user's friends or pages they follow, or come from sources that post about topics that the user seems to be interested in. The posts can also be curated by the platform, such as information about voting locations during an election cycle or about COVID vaccines.  

Because much of the content is organic, the first step in the process is to remove items that violate the community policies set forth by the platform. Such items may include hate speech, bullying, nudity, calls for violence, or misinformation~\cite{DBLP:journals/corr/abs-2009-10311}. 
At the next step, the recommender uses a lightweight model to reduce the number of candidate items to about 500 per user. 

In the main ranking step, the goal of the recommender system is to produce a score $c_{p,u,T}$ which determines whether post $p$ is of value to user $u$ at time $T$. The value of $c_{p,u,T}$ is computed as a weighted linear sum of a set of prediction models. Prediction models can include trying to predict engagement actions such as liking or resharing a post. They can also incorporate  how recent the post is or whether other friends have engaged with it. Since not all types of user value can be captured through their engagement with posts, the platform also conducts surveys with small sets of users. These surveys can ask direct questions such as whether posts are worth their time, provide value, or contribute to their feeling of being informed. Based on these surveys, an additional set of predictors tries to predict the user's answer to these survey questions. 

The final step of the recommendation system is a re-ranking step, where the goal is to ensure that the feed that is produced contains enough diversity (e.g., not overwhelmed by a particular topic or set of users that post often). 

\subsection{Affective response}
\label{section:af}

In this work, we use the term \textbf{affect} to refer to both an affective state (e.g., anger, joy) as well as a cognitive state (e.g., entertainment, inspiration). Our goal is to model the \textbf{affective response}, the affect that a user may have when exposed to a particular post, and use such a model to improve a scoring module of the recommendation system. 

Predicting affective response is a form of content analysis. 
A common type of content analysis done for recommendation today is to determine the {\em topic} of a post \cite{bergamaschi2014comparing, pennacchiotti2011investigating}. However, the topic alone does not capture its full nuance. For example, consider that a post on the topic of clowns can have an affective response varying from happy to scared. Our definition of affect as both affective and cognitive experiences is motivated by a need to narrow the gap between topic analysis and more nuanced analysis of a post.

We found that a prediction of affective signals can be useful in several of the scoring modules that are used in the recommender system, and therefore we create an embedding that can be used as a feature in any scoring module. The embedding is the penultimate layer of the network that predicts affective tasks.

An alternative option would have been to use the affective model as a scoring module in itself. We discarded this option because these affective signals alone do not necessarily indicate whether a post should be ranked higher or lower.

Finally, it is important to differentiate between affective response and the more prevalent work in Affective Computing on emotion detection. In the context of social media, emotion detection has been used to try to detect the emotion of the author of the post (referred to as the \textbf{publisher affect} in~\citet{chen2014predicting}), whereas our focus is on the anticipating the viewer's affective response to the post. While the publisher affect may be relevant to the affective response, it is not always sufficient signal (see Figure~\ref{fig:clown} for an illustrative example). For example,  a post with an excited publisher affect can induce an angry affective response. We decided to focus on affective response for two main reasons. First, detecting the emotional state of the author of a post is ethically questionable. Second, our goal here is to curate content that caters to the preferences of the user, and therefore the possible affective responses should be taken into consideration.
 
\subsection{Defining the affective taxonomy}
\label{section:taxonomy}


In determining the set of affective responses to operationalize, we need to construct a taxonomy of labels with the proper granularity. There are two challenges in designing the taxonomy. First, the labels need to be discriminative enough to distinguish different use cases and serve a variety of user preferences. Take, for instance, the affective response of being \textit{angered}. On social media, there are instances where collectively venting over a common issue can permit self-expression and community, and can be cathartic and useful to both the posters and viewers \cite{vermeulen2018smiling, jalonen2014social}. On the other hand, there is clearly an unproductive type of anger that can arise, such as when viewing posts containing spam, toxic speech, or misinformation~\cite{cinelli2021dynamics}. In the former case, we may still consider showing the user the candidate post, but in the latter case it's much more unlikely. Consequently, we realized that a single label for \textit{angered} was too broad, and we constructed two types of \textit{angered}: \textit{constructively-angered} and \textit{deconstructively-angered}. Another example is excited and relaxed, which in other works such as \cite{shaver1987emotion} are classified under a single category \textit{joy}.

The second challenge we faced was to design a taxonomy that covers the important use cases but, for practical considerations, also minimizes the number of affective responses as much as possible. As noted in~\cite{karger2013efficient},  the degree of difficulty in modeling affective responses would scale with an increase in the number of affects. In particular, the quality of human labels would decrease because it is impractical to ask annotators to distinguish between a large number of affects. In Section~\ref{section:evaluation}, we evaluate the interrater correlation and show average values much larger than that of state-of-the-art work in publisher affect detection.


\begin{table}[htb]
  \small
  \caption{Affective responses (class) and their corresponding definitions of our taxonomy.}
  \label{tab:taxonomy}
  \begin{tabular}{l l }
    \toprule
    Class & Definition\\
    \midrule
    Adoring & Response to finding something adorable. \\
Connected & Feeling more connected (either to the person \\
& making the post or something in the post).\\
Constructively- & Angered in response to content that is angering\\
angered & but informative, valuable, or promoting social good.\\
Destructively- & Angered or annoyed in response to unproductive,\\
angered & unhealthy, borderline violating content.\\
Entertained & Finding entertaining, amusing, or humorous. \\
Excited & Feeling joy, excitement, enthusiasm or eagerness.\\
Grateful & Feeling grateful or appreciative.\\
Informed & Feeling of being informed or having received new \\
& information regarding a subject, event, or topic.\\
Inspired & Feeling inspired, motivated, uplifted, or encouraged.\\
Neutral & Having a neutral feeling.\\
Relaxed & Feeling peaceful, calm, or relieved.\\
Saddened & Feeling grief, unhappy, sad.\\
Scared & Feeling of concern, anxiety, fear, or stress.\\
Surprised & Feeling shocked or astonished (either +/-).\\
Touched & Feeling moved, emotionally stirred.\\
  \bottomrule
     \end{tabular}
\end{table}

\section{Methods}
\label{section:methods}



In this section, we describe methods for generating training data for our affective models and the architecture and features of the model that is trained on this data. Training data can be generated in two ways: engagement data that we have on the platform and human labeling of posts. Our goal is to leverage the engagement data as much as possible to reduce the costs of human labeling. Section~\ref{section:engagement} describes how we extract training data from these simple behaviors on the network (e.g., reactions to post). Note, we could have also relied on explicitly asking users their affective states regarding content, but this has the drawbacks of being intrusive and potentially unreliable~\cite{pantic2009implicit}. In Section~\ref{section:comments} we analyze the content of comments written in response to posts to generate training labels, and in Section~\ref{section:pdo} we discuss how we obtained annotations from human labelers. Together, the affective response labels and the engagement labels are used for training a two-tower architecture multi-class classifier described in Section~\ref{section:model}. For this work, only de-identified Facebook posts were utilized. 

\subsection{Engagement signals}
\label{section:engagement}


When using the platform, there are a number of ways that users can indirectly give feedback on how they feel about the content that they see. These include but are not necessarily limited to: 

\begin{itemize}
\item User reactions (e.g. like, love, care, haha, wow, sad, angry as in Figure~\ref{fig:reactions})
\item User behaviors (e.g. share, outbound click)
\item Negative user feedback (e.g. hide, snooze, unfollow, report as in Figure~\ref{fig:nuf})
\end{itemize}

While these signals may not be directly indicative of an affective state, a subset of these categories could conceivably provide useful signals that are transferrable to learning the affective response. For instance, a user clicking the haha reaction to a post might indicate that they're feeling entertained by that content, while a user reporting a post might indicate that they are angered or offended. These engagement signals are also straightforward to incorporate into our model because they are already personalized (i.e., the engagement signal is both user and post dependent). 

In the dataset we constructed, we used prediction of these engagement signals from the prior 90 days as a new training task (i.e., as prediction labels). Another possibility would be to use these engagement signals as features, rather than training labels, because they intuitively could help in predicting the affective response. There are two issues with this. Firstly, removing engagement as training labels, would greatly reduce our overall training set size and force us to learn a more complex feature space with less data. Secondly, we ideally would like to predict affective response at the time of the post's creation when there is little to no engagement yet. Consequently, assuming engagement signals are available to be used as features was not a suitable design.

\subsection{Patterns from comments}
\label{section:comments}

In addition to engagement, comments that users write in response to posts contain valuable signal relevant to the affective response they had when viewing the post. For example, the expression ``What a hilarious story'' may indicate that a post is humorous, and ``This is so cute'' may indicate that a post is adorable. If comments of the same flavor appear multiple times in a response to a post, we can use that as a label for the affective response.

We developed the CARE (Common Affective Response Expression) method~\cite{yu2022care}, a means of obtaining labels for affective response in an unsupervised way from the comments written in response to online posts. Since these labels were going to be used as training data, we wanted to ensure that their precision is high, but we also wanted a flexible method that can be applied for new affects as they came up. 

\begin{figure*}
  \centering
  \includegraphics[width=0.78\linewidth]{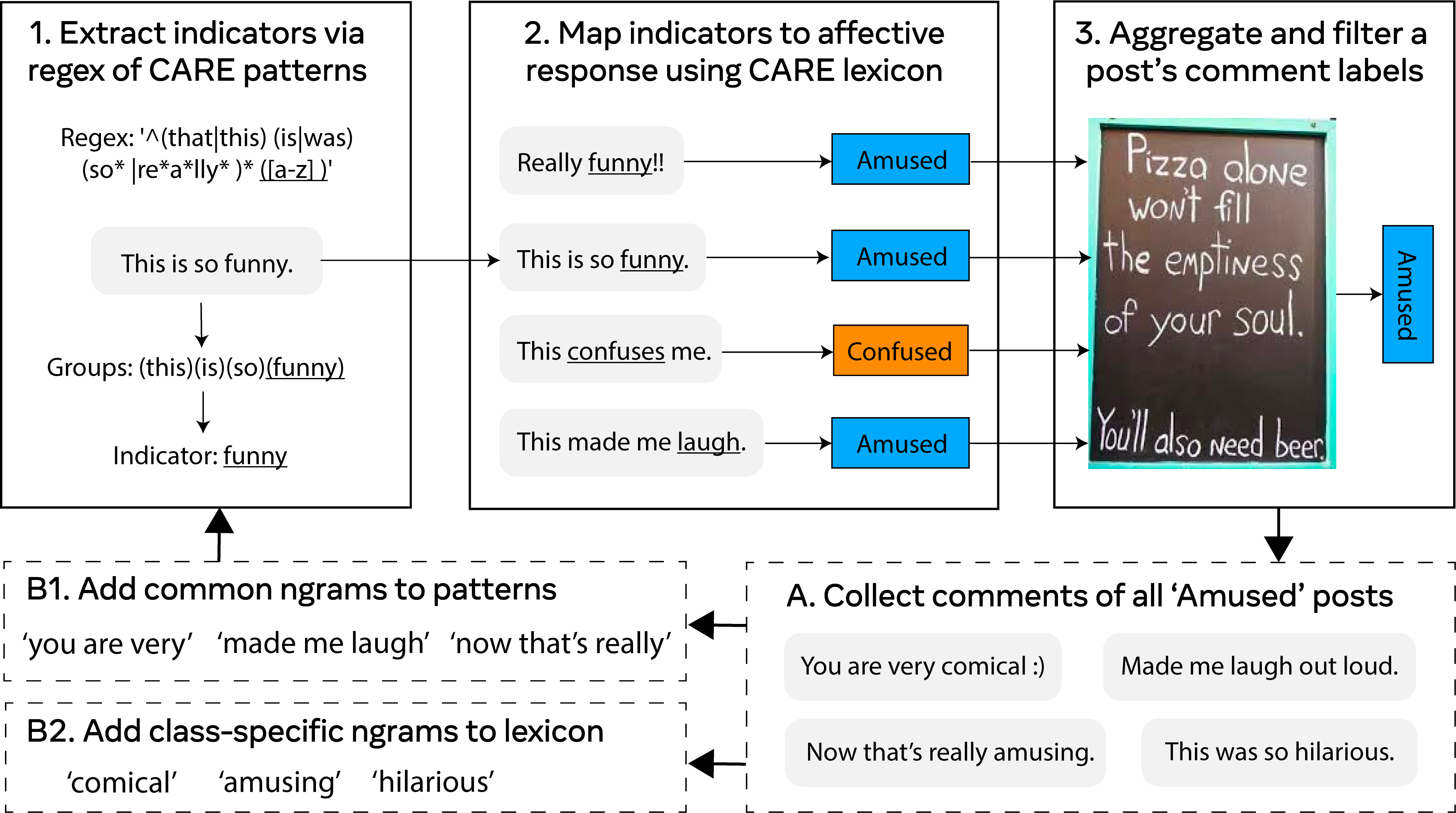} 
  \caption{Overview of the CARE Method. The top part of the figure shows the process of labeling a post, while the bottom features how we expand the set of patterns and lexicon. In step (1), we apply CARE patterns to the comment and extract the indicator or keyword. In step (2), we map each comment to the corresponding affective response using the CARE lexicon. Step (3) aggregates the comment-level labels for a post-level label. In step (A) we collect all comments of all posts corresponding to a particular affective response and analyze its most frequent n-grams. N-grams common to multiple classes are added to the CARE patterns (B1) while frequent n-grams specific to a class are added to the lexicon (B2). This process can then be repeated until a satisfactory number of labels are generated.}
  \label{fig:care_diagram}
\end{figure*}

CARE uses patterns and a keyword-affect mapping to identify expressions in comments that provide high-precision evidence about the affective response of the readers to the post.  We seed the system with a small number of high-precision patterns and mappings. We then iteratively and automatically expand on the initial set by considering frequent patterns and keywords in unlabeled comments on posts labeled by the previous iteration. The CARE method is illustrated in Figure~\ref{fig:care_diagram}. We stopped expansion of the system after reaching 23 distinct patterns and a lexicon of 163 keywords because this sufficed to generate enough labels for each class. Note, because these patterns are applied to the comments rather than the post, training any models on the content of the post will not be biased to these particular expressions.

Using the CARE method, we obtained 4 million labels for the affects \textit{adoring}, \textit{entertained}, \textit{excited}, \textit{saddened}, \textit{scared}, \textit{angered}, and \textit{approving} (which is not in Table~\ref{tab:taxonomy} but refers to expressing support, praise, or pride). To evaluate the quality of the labels generated by CARE, we randomly selected 6000 posts and asked human annotators to label them according to these labels. The comparison of the human annotations and the CARE labels are shown in Table~\ref{tab:amt_results} and indicate high agreement.

In implementing CARE, we noticed a few shortcomings of the method. Firstly, in analyzing the individual classes and mappings, there are some patterns which work for certain class-keyword mappings but not so for others. While this is a point for improvement in future work, we observe that in large numbers, the more error-prone combinations are infrequent compared to the highly-accurate ones. Consequently, the method as a whole is a reasonable cost-effective alternative to obtaining more human annotations. 

\begin{table}[htb]
\small
\centering
\begin{tabular}{cccc} 
\toprule
 \textbf{\# Agree} & \textbf{Any CARE} & \textbf{All CARE} & \textbf{Other} \\
\midrule
  $\ge 1$ & 98 & 96 & 82\\
  $\ge 2$ & 94 & 90 & 53\\
  = 3 & 80 & 76 & 24\\
  \bottomrule
  \hline
\end{tabular}
\caption{The rate of agreement between the annotators and the labels proposed by CARE.  The first column specifies the number of annotators to be used for consensus. The rest of the columns shows for all posts, the average rate of intersection of the human labels with at least one CARE label, all CARE labels, and any label that is not a CARE label.}
\label{tab:amt_results}
\end{table}

\subsection{Human labels}
\label{section:pdo}

In addition to weakly-labeled data from engagement signals and comments, we also crowdsourced a ground-truth dataset, which 
allowed us to augment our original data with high-quality labels that can be used for evaluation and analysis. In our annotation tasks, we restricted to posts with just text and image (no video) and to English only. 

\paragraph{Labeling guidelines} 
Before deciding the details of the annotation procedure, it was important for us to first understand which frame of reference we wanted annotators to label from. Unlike in most labeling frameworks which ask objective tasks (e.g., is the post about baseball?), a labeler's background and personality can greatly affect their answers in our context. More specifically, in asking individuals about the affective response to a post, they could either answer with their personal opinion or they could answer with what they perceive is a more universally accepted answer:
\begin{enumerate}
    \item\label{itm:personalized_pdo} Personalized: The affective response of a post from a labeler's personal perspective (i.e., how does this post make {\em you} feel?).
    \item \label{itm:unpersonalized_pdo} Unpersonalized: The affective response of a post from a common, universal perspective (i.e., how does this post make {\em most people} feel?). One could equate this framing to how genres on Netflix are labeled (e.g., feel-good, emotional, provocative).
\end{enumerate}

It is clear that (\ref{itm:personalized_pdo}) lends itself best to personalized predictions and is more akin to the recommendation system setting. However human annotators are not our users and therefore we lack their user engagement history and other information that is normally critical for personalized prediction. Moreover, for each post, we have five distinct labelers annotate, which is not sufficiently large  to leverage annotator agreement if operating under (\ref{itm:personalized_pdo}). For these reasons,  (\ref{itm:unpersonalized_pdo}) was the more effective solution, and so we asked the following question: How might someone feel after seeing the following post? Select up to 3 of the top options.


\paragraph{Personalizing annotated posts} 

As discussed previously, the labels we obtained from human annotation were not personalized, but our recommendation system is personalized. In order to incorporate the non-personalized human labels, we construct a personalized dataset with the following heuristic: if a user liked (or loved) a post that was annotated with an affect $A$ by the annotators, we assume that the user also had the same affect towards the post. Specifically, for each post $p$ labeled as a positive affect $A$ by the human labelers, we collect all users $\mathcal U$ who liked or loved the post. We then add a row $(p, A, u)$ to the personalized dataset for each $u \in {\mathcal U}$. We note that this heuristic does not apply to the negative affects (\textit{angered}, \textit{saddened}, and \textit{scared}), since `liking' an angering post does not naturally imply that the user felt angered. Personalizing these negative affects remains as future work and they were withheld from the modeling stage.  This personalization process also applied to the CARE labels discussed in Section~\ref{section:comments} but was not necessary for the engagement-sourced labels since those are already personalized. 



\subsection{Modeling}
\label{section:model} 

Now that we have discussed each component of the training data, we combine all engagement signals (all reactions, unfollow, report, hide, share, and outbound click) and all affective response classes except \textit{angered}, \textit{saddened}, \textit{neutral}, \textit{other}, and \textit{scared} (excluded due to the reasons discussed in the previous paragraph), resulting in a total of 23 classes. The training process involves taking a subset of 1 million samples from each prediction class, where half are positive samples and half are negative samples. We create a train/validation/test split of 80/10/10, respectively, on a dataset totaling 23 million samples. 

\begin{figure}[hbt]
  \centering
  \includegraphics[width=\linewidth]{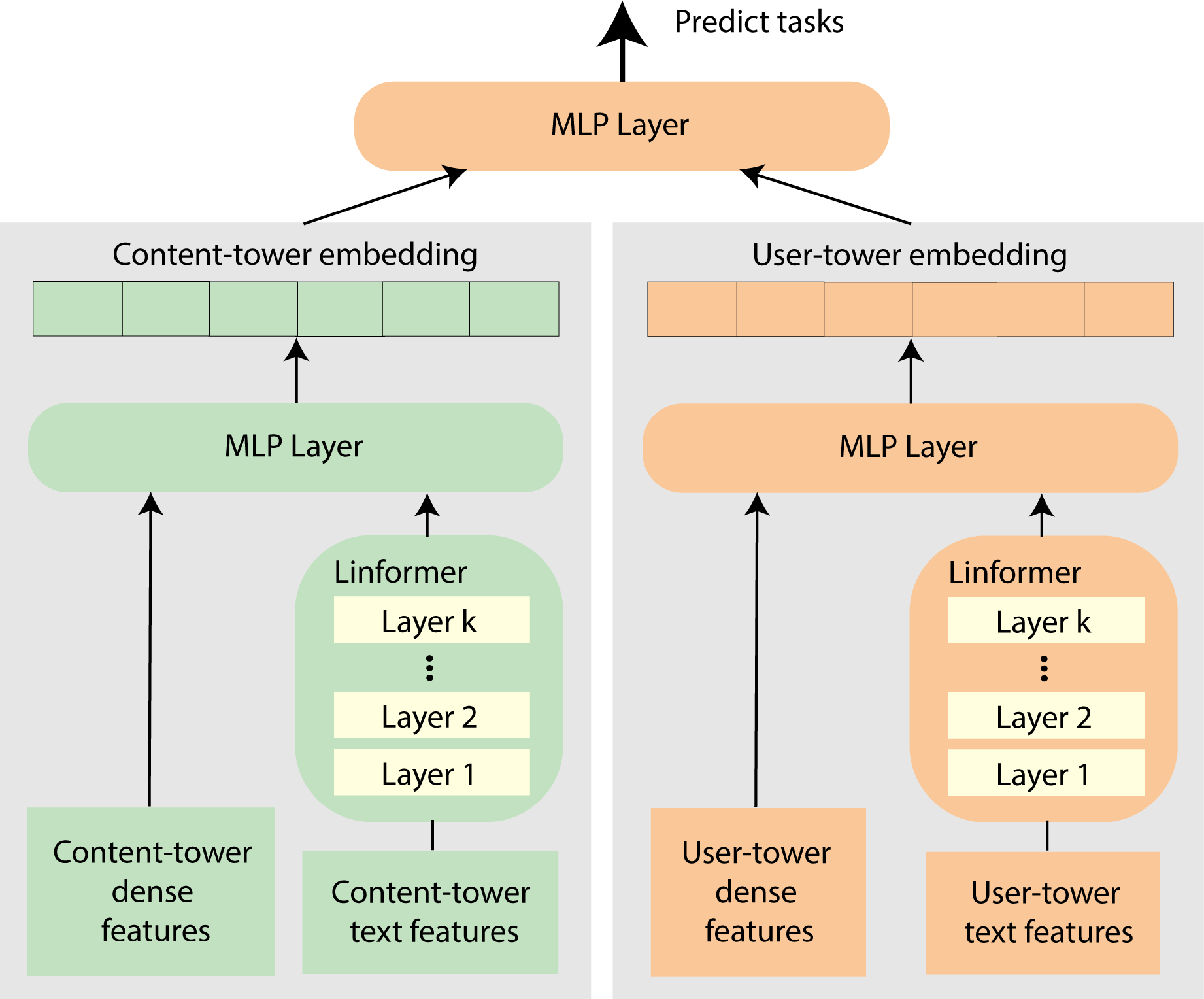} 
  \caption{Architecture of the two-tower model. The left and right tower encode content and user information, respectively. The model is trained using the engagement signals and the affective response labels described in Section~\ref{section:methods}. Each tower encodes text features and dense features, where dense features are typically embeddings from other models which may capture high-dimensional information. One such embedding would be one that describes the visual components of an image in the post.}
  \label{fig:pps_arch}
\end{figure}

Conceivably, our multi-label classification model trained on affective signals can be used for a number of applications. In some uses cases, personalization will be necessary, and in others, only content information be necessary. In order to have a model which is flexible to the needs of the downstream use cases, we used a two-tower architecture model (see Figure~\ref{fig:pps_arch}) where the left tower is used to model information about the content, and the right for the user. Each tower features a linformer transformer \cite{wang2020linformer} followed by a multi-layer perceptron (MLP) module. The outputs of each tower are then fused in a secondary MLP module, which is then used for multi-label classification. 

More concretely, the input to the model consists of the features for both the content and user tower. The content features, for instance, consist of properties of the post like the text of the title, body, optical character recognition, and video transcript, if available. The user features, on the other hand, consist of statistics from the user's network, interests, and profile properties such as text from a user's biography. The output of the model is a multi-label prediction for the 23 prediction classes (i.e., a vector of length 23 with binary values). After conducting a sweep over epochs and learning rate, we found training with 3 epochs and a learning rate of 0.0007 to be optimal.

\section{Results}
\label{section:evaluation}

In this section, we provide results and analysis of the dataset and model. Specifically, we first discuss statistics pertaining to the human-labeled dataset (e.g., annotator agreement) and second, we will discuss correlation with alternative labels, such as those discussed in Section~\ref{section:engagement}. Lastly, we describe results for incorporating the trained model into the overarching recommendation system. 

\subsection{Analysis of human annotations}

\begin{figure}[hbt]\vspace{1em}
  \centering
  \includegraphics[width=\linewidth]{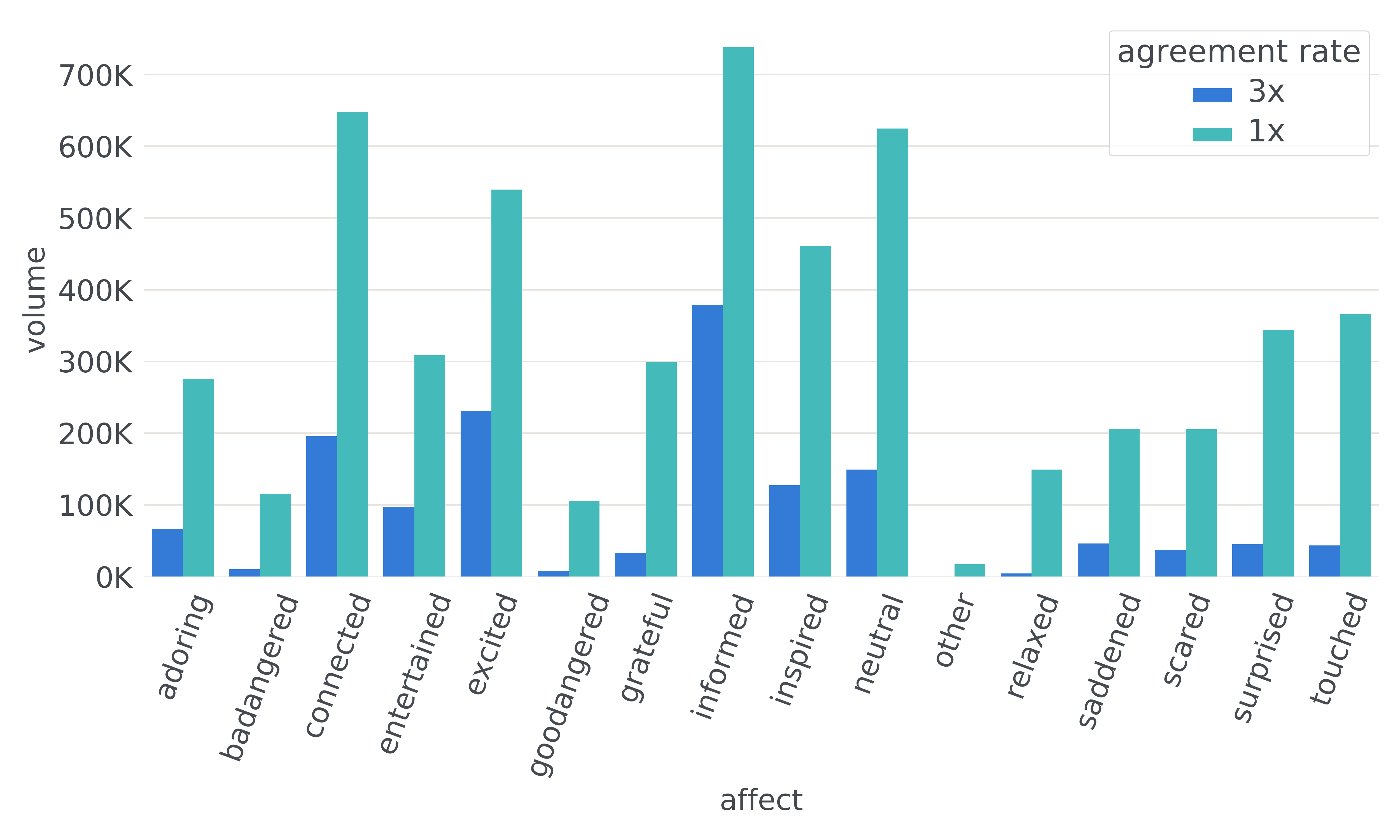} 
  \caption{Number of annotations per affect where at least 3 out of 5 annotators agree (3x) or where any annotator selects the label (1x).}
  \label{fig:pdo_volume}
\end{figure}

We collected human labels for nearly 820k posts with five annotators each, resulting in a total of 7.3 million annotations. In total, there were 348 unique human annotators and on average, each annotator selected 2.57 options per post. In analyzing the number of annotations per post, we compute these statistics using two sets of labels: labels selected by at least one out of the five annotators (1x) and labels selected by at least three of the five annotators (3x). Figure~\ref{fig:pdo_volume} shows the breakdown by class under both settings and indicates that the labels \textit{informed}, \textit{excited}, and \textit{connected} are among the top most prevalent affects. Note, we did not source posts randomly. Instead, we iteratively trained simple binary models to predict for each affect, applied inference to a set of randomly selected posts, and then sent posts with high prediction scores (particularly for low-volume and important classes such as \textit{entertained} and \textit{inspired}). This was done in an effort to screen out neutral posts and reduce labeling costs. Thus, the distribution in Figure~\ref{fig:pdo_volume} is not reflective of the sampling distribution. 

\begin{figure}
  \centering
   \includegraphics[width=\linewidth]{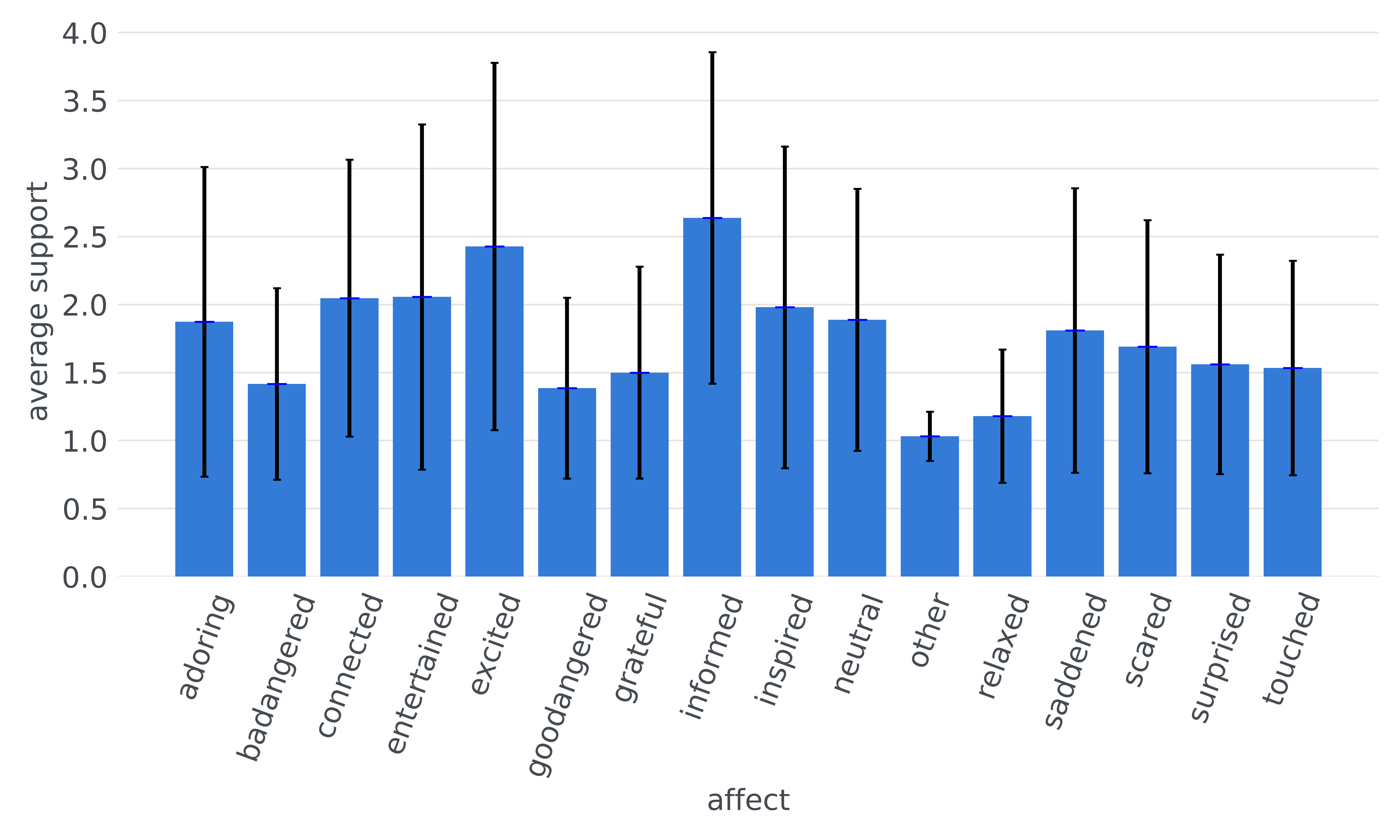} 
  \caption{Average degree of annotator support for each class. Error bars indicate standard deviation across all posts.}
  \label{fig:annotation_support}
\end{figure}

Naturally, the number of labels agreed upon by 3 annotators (3x) is smaller than labels with no agreement restriction (1x). The average degree of annotator support for each affect (given that at least one annotator suggests the affect) is shown in Figure~\ref{fig:annotation_support}. The numbers suggest that the classes \textit{other}, \textit{relaxed}, the two types of \textit{angered}, and \textit{grateful} have the lowest degree of agreement given 5 annotators. The \textit{other} category here was an option in the labeling process to submit alternative affects not listed in our taxonomy. Some of the most frequent suggestions were \textit{confused}, \textit{curious}, \textit{yummy}, \textit{beautiful}, \textit{disgusted}, and \textit{annoyed}, and these alternative suggestions are valuable for improving future iterations. We did consider distinguishing \textit{disgusted} and \textit{annoyed} from \textit{destructively-angered}, but concluded that the use cases are too similar to justify additional labels. 

Following the work of \citet{demszky-etal-2020-goemotions} on publisher affect, we estimate rater agreement by interrater correlation~\cite{delgado2019cohen}, which is computed by taking the average correlation between each rater's judgement and the mean of other rater judgements. We find that the average inter-rater correlation in our context is 0.52, which is much higher than the inter-rater agreement of 0.28 in~\cite{demszky-etal-2020-goemotions}, where they had 28 classes in their taxonomy (and only~3 labelers).   



\begin{figure}
 \centering
  \includegraphics[width=\linewidth]{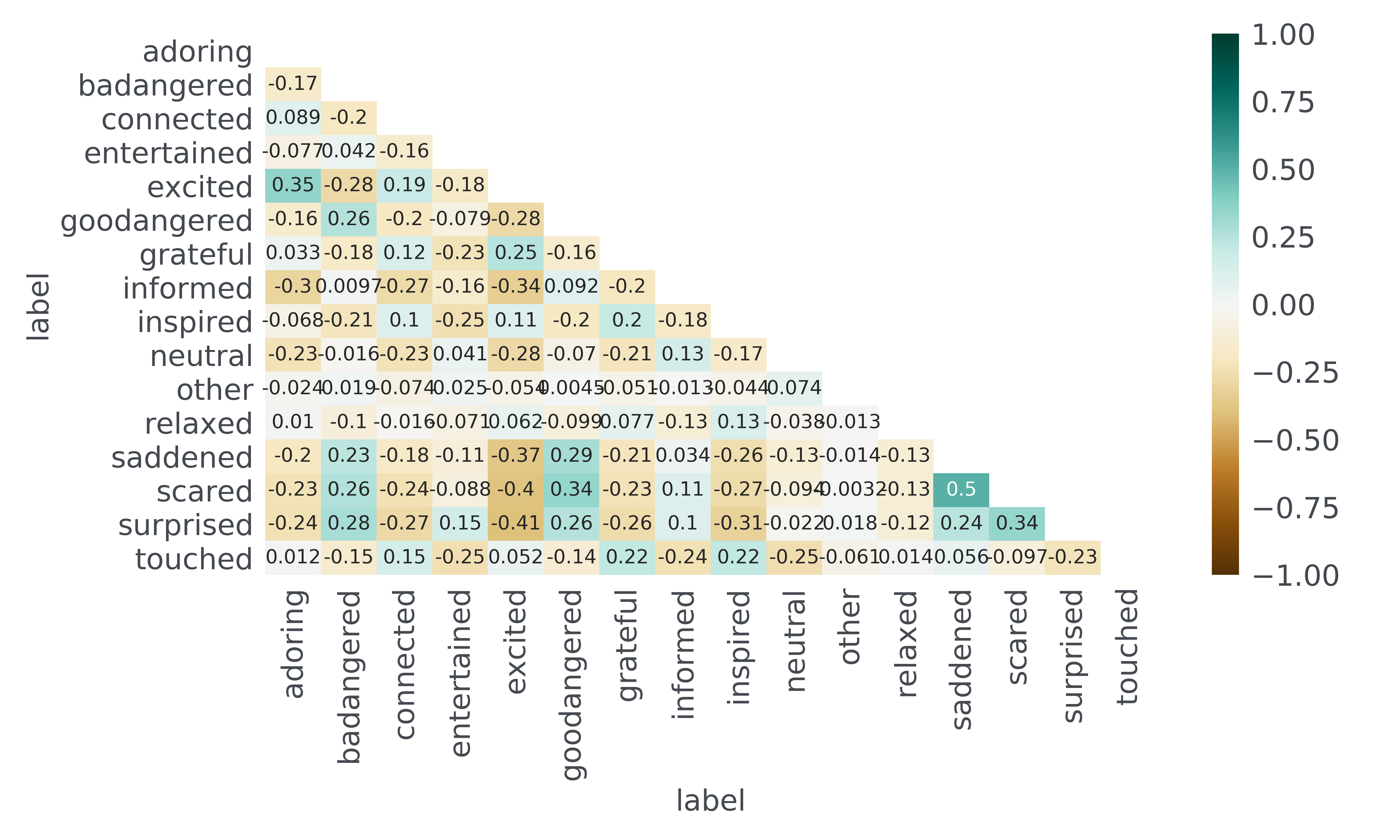} 
  \caption{Pearson correlation between the different affective responses in the taxonomy.}
  \label{fig:affect_correlation}
\end{figure}


To investigate whether the low agreement of certain classes is a consequence of conceptual overlap amongst the labels, we study the correlation between the affects in Figure~\ref{fig:affect_correlation}. Here we use the interpretation of moderate (0.40--0.69) and weak (0.10--0.39) correlation as described in \citet{schober2018correlation}. Figure~\ref{fig:affect_correlation} shows that weak correlations are indeed present. For instance, \textit{constructively-angered} and \textit{deconstructively-angered} are weakly correlated with each other (0.26) as well as several other affects like \textit{scared}, \textit{saddened}, and \textit{surprised} (0.23 to 0.34). Other correlations that are weak to moderate include \textit{adoring} and \textit{excited} (0.35) as well as \textit{saddened} and \textit{scared} (0.5). Interestingly, the  affective response \textit{connected}, which is not typically in traditional emotion detection taxonomies, seems to have weak correlations with feelings of excitement (0.19), gratitude (0.12), inspiration (0.1), and feeling touched (0.15). 

\subsection{Publisher affect vs.\ affective response}

In Section~\ref{section:taxonomy} we distinguished between publisher affect of the poster and the affective response of the viewer, noting that the two are not always interchangeable. In what follows, we experimentally validate and quantify this hypothesis.

Like other social networking platforms, there exists a feature allowing posters to set their status updates, particularly with feelings such as \textit{feeling blessed} or \textit{feeling sad}, which we refer to as the poster-annotated feelings. Because the poster-annotated feelings come from the users themselves and denote their affective state, we can consider these labels as the publisher affect of the post. Hence, we can test our hypothesis given these two sets of labels: publisher affect from the user and affective response from human annotation. 

For this analysis, we identify posts in our human-annotated dataset which contain poster-annotated feelings and compute their correlation (after filtering out feelings with frequency less than 1000). As shown in Figure~\ref{fig:minutiae_correlation}, we find moderate correlations between \textit{feeling sad} and \textit{saddened} (0.5) and weak correlation for \textit{feeling worried} and \textit{scared} (0.36), which are intuitive relationships. Overall, however, these correlations between equivalent publisher affects and affective response such as for \textit{excited} (0.12) and \textit{grateful} (0.11) are weak, suggesting that making a distinction between publisher affect and affective response is valid. Interestingly, \textit{touched} is weakly correlated with \textit{feeling sad} (0.25) and \textit{deconstructively-angered} and \textit{constructively-angered} are weakly correlated with \textit{feeling angry} and \textit{feeling annoyed} (0.19 to 0.25), though this is more so the case for \textit{deconstructively-angered}. While there is certainly noise in making these comparisons, the low degree to which these two types of affects correlate indicate that they are not completely synonymous.

\subsection{Affective response vs.\ engagement}

While the poster-annotated feelings give insight into the poster's perspective, we also want to understand how affective response aligns with the viewer's actual behavior and feedback.  Figure~\ref{fig:affect_engagement} shows the correlation between several prevalent engagement signals and affective response labels. Firstly, the behavior and negative user feedback signals (nufs) don't seem to have significant correlation with any of the affects ($<$ 0.042), but this is perhaps due to their low overall prevalence (less than 1000). For reactions, where data is more abundant, we find that the anger, haha, and sorry reactions are weakly correlated with the affective responses \textit{angered} (0.14 to 0.25) and \textit{entertained} (0.41), and \textit{saddened} (0.36), respectively, as one might anticipate. Intuitively, like and love reactions are associated with positive affects  and inversely so to negative affects. The support reaction seems to be correlated most with \textit{saddened} (0.14), which empirically seems to be because it is often used in response to expressing concern or sympathy to sad news. As seen earlier in Figure~\ref{fig:affect_correlation}, the anger reaction is also weakly correlated with \textit{saddened} (0.13), \textit{scared} (0.16), and \textit{surprised} (0.11).  These results together suggest that some affective signals can be gleaned from engagement, as discussed in Section ~\ref{section:model}.

\subsection{The affective model in the recommender}

This section provides an analysis of the end-to-end system that incorporates the affective predictors.  To evaluate our affective model, we experimented with using our model embedding in one of the scoring models of the recommendation system. Given a post and a user, the scoring model we chose tries to predict a user's answer to a survey concerning their preferences regarding the post. We chose this predictor for two reasons: (a) understanding whether a user wants to see more of a particular content necessitates understanding the user's affective response to a given post and (b) since this scoring model tries to predict the result of a survey, it by nature has much less training data, and hence can potentially benefit from the data-rich affect embeddings. 


We conducted several ablation experiments, which involved a two-step process. The first step involved ablating features and parameters such as the number of encoding layers. The second step involved exporting the 32-length embedding from the content tower, and using this as a feature in the scoring model. Here we prefer to use the embedding from the content tower for internal infrastructure efficiency reasons, but we experimented with both the content and user tower embeddings (including the concatenation of both) and found the results for the latter to be only marginally higher than the content embedding alone.

To evaluate, we first created a static dataset of around 1.5 million survey responses. After running ablation experiments using this offline dataset that was split into train, validation, and test partitions, we identified an embedding with the highest statistical improvement. This model achieved an AUC-ROC loss reduction (the observed improvement normalized by the possible amount of improvement) of more than 8\%, as computed by $\frac{S_{new} - S_{base}}{1-S_{base}} *100$ where $S_{new}$ and $S_{base}$ refer to the AUC-ROC of the model with the embedding and model without the embedding, respectively. We then conducted an online experiment involving more than 20 million users using the scoring model which uses the new affective embedding (the test group) and using the original scoring model running in production, which does not use the affective embedding (the control group). During the 14 days of online experimentation, we measured a number of metrics relevant to user satisfaction with the platform and benchmarked these values against those of the control group. 

The results showed statistically significant decreases in visibility of integrity-violating content like misinformation and engagement bait (more than 0.6\% decreases), and also demonstrated meaningful gains in engagement, like the number of like reactions (around 0.4\% improvement), without causing detriment to other key important metrics. 
Additionally, the affective embedding ranked as the most important feature in the scoring model, reaffirming our approach to ranking from an affective response perspective. It also suggests that users, when given more control over their content, will overall choose higher quality content that encourages greater engagement. After launching to an even larger population, these overall trends still generally hold.

\section{Related work}
\label{section:related}

In this section, we situate our work with respect to previous research on related tasks.

\subsection{Affective Recommender Systems}

Our work is the first to demonstrate that affective signals can benefit recommender systems at large scale in the context of social networks. The following are related work in the field. A significant portion of these study implicit physiological signals like facial and audio tracking \cite{vrochidis2011utilizing, soleymani2011multimodal, kierkels2009queries, tkalcic2012affective} while others rely on explicit surveys \cite{lang2005international, gonzalez2007embedding, tkalvcivc2010using}, both of which are not feasible for social media at scale. Some of these works also focus on clean and curated multi-media datasets that are impractical for real-world settings \cite{odic2012relevant, odic2013predicting, garcia2013pessimists, zheng2013role}. \citet{orellana2015mining} applies affective recommendation to social media, particularly Youtube videos, by acquiring affective annotations from 80 human annotators along Plutchik’s eight basic emotions \cite{plutchik1980general}. Much larger in scale is the work done by \citet{mizgajski2019affective} and \citet{leung2020text} using feedback to online news and tweets, respectively, but these works do not leverage multiple types of affective sources. \citet{qian2019ears} combines user rating data, user social network data, and sentiment from user reviews as affective information, but again is small in scale. Additionally, many of these works utilize traditional machine learning techniques like similarity-based clustering or regression trees as the basis for their recommendation system \citep{mizgajski2019affective, orellana2015mining}. We note that this work also differs from others in the space of affective models in that it utilizes a two-tower architecture to jointly model the user and content features, particularly for live online prediction serving billions of users. 

\subsection{Methods for unsupervised labeling}
\label{sec:weak_labeling}

A major bottleneck in developing models for emotion and affective response detection is the need for large amounts of training data. As an alternative to manually-labeled data, many works utilize metadata such as hashtags, emoticons, and Facebook reactions as pseudo-labels~\cite{wang2012harnessing, suttles2013distant, hasan2014using, mohammad2015using}. The work we present here extracts labels from both engagement like Facebook reactions as well as free-form text in comments rather than metadata. The work done in~\citet{sintsova2016dystemo} is similar to our work on comments in that it pseudo-labels tweets and extends its lexicon, but the classifier itself is a keyword, rule-based approach and is heavily reliant on the capacity of these lexicons. In contrast, our work leverages the high precision of CARE on the comments and uses the post content to train a model, which is not constrained by the lexicon size in its predictions. Our method also employs bootstrapping to expand the set of patterns and lexicon, similar to \citet{agichtein2000snowball} and \citet{jones1999bootstrapping} but focuses on extracting affect rather than relation tuples. Many works utilize engagement and social network structure as features instead of labels in their model \cite{lipczak2012understanding, he2014predicting, yang2013sentiment, qian2019ears}, but as explained in Section~\ref{section:engagement}, our application needs to perform inference prior to engagement signals being available.

\subsection{Affective taxonomies}

Perhaps two of the most well known categorical organizations for emotion are Paul Ekman's six basic emotions (happiness, sadness, disgust, fear, surprise, and anger) \cite{ekman1999basic} and Robert Plutchik's Wheel of Emotions (anger, anticipation, joy, trust, fear, surprise, sadness, and disgust) \cite{plutchik1980general}. Arguably all of Ekman's six basic emotions exist in our taxonomy, with the exception of disgust which is assumed by deconstrutively-angered. These basic emotions are hardly sufficient, which is in line with Plutchik's theory that suggests few experiences are basic ones---they are often combination results, which necessitates the need for a more comprehensive taxonomy in practice. The Flickr LDL dataset \cite{yang2017learning}, for example, contains images labeled according to a taxonomy that uses Ekman's six but also includes amusement and contentment, akin to entertained and relaxed in our current work. We also know from prior work that adequately detecting inspirational \cite{ignat2021detecting} and informative \cite{ni2007exploring, malla2021covid} content as well as content expressing gratitude \cite{sciara2021gratitude} is beneficial for users. \citet{craig2021can} also found that the primary reasons adolescents use social media is because they want to be entertained, be informed, and feel connected to others. Our taxonomy builds upon prior taxonomies, but includes affects with the intention to satisfy these user needs.


\section{Conclusions}
\label{section:conclusions}

We described the challenges involved with incorporating affective signals in a large-scale recommendation system and the solutions we developed at Facebook. In particular, we designed an affective taxonomy  customized to user needs on social media, and created training data for our models by combining engagement data and data from a human-labeling task. Our two-tower model learns from both engagement signals and affective response labels. Our results also  provide new insights into the correlations among the affects in the taxonomy and correlations between publisher affect and viewer affective response, thereby justifying some of the design choices we made.   We demonstrated that exporting the embedding of this model and using it as feature in one of the scoring models of the recommendation system greatly improves performance, both online and offline.

There are several avenues for improvement and additional research. Our taxonomy can be extended with  affective responses that the human annotators  frequently noted as missing and our techniques for personalizing human labels need to be extended to negative affects. We believe that more advanced analysis of images and videos can improve our models considerably. More broadly, our work considered the affective response the user may have to a {\em single} post. However, it is not clear how these individual affective responses combine to an affective response for a session that includes a {\em sequence} of posts, which is closer to the overall experience the user has on the platform.

\bibliography{main}
\bibliographystyle{acl_natlib}

\clearpage

\appendix

\section{Broader Impact}

Any work that touches upon recognizing affective response needs to ensure that it is sensitive to its application. Our work in detecting affective response is intended for anticipating the affective response of users to content, in order to better safeguard them against offensive material and provide them with content that better aligns with their user preferences. This work should not be used for ill-intended purposes, such as purposefully recommending particular content to manipulate a user's perception or preferences. Additionally, any work that utilizes user information or content created by users must be careful in respecting the privacy preferences of its users. Before this research was conducted, it went through an extensive internal review process with a diverse team to delineate these bounds. Regarding our crowdsourcing process, the human annotators were paid a competitive and fair rate. The raters were also selected by diversifying the pool amongst several categories along five attributes: age, ideology, gender, ethnicity, and location.

\section{Affective Response vs. Publisher Affect}

\begin{figure}[H]
\centering
  \includegraphics[width=\linewidth]{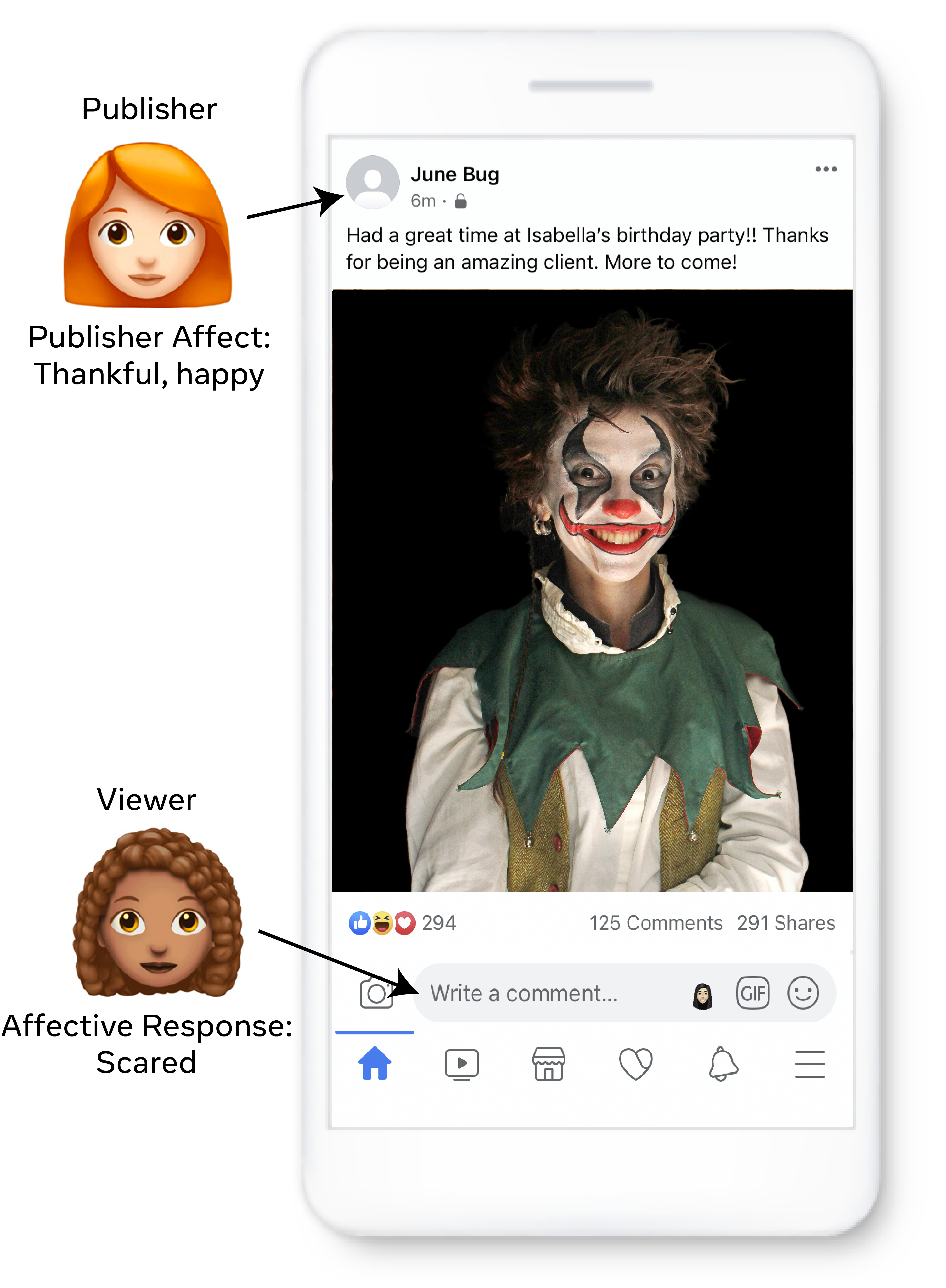}
  \caption{An example case of differing publisher affect and affective response. This work focuses on affective response through signals such as comments and reactions. Post image sourced from Shutterstock \cite{tapia}.}
  \label{fig:clown}
\end{figure}

\section{Facebook Interface}

\begin{figure}[H]
\centering
  \includegraphics[width=0.55\linewidth]{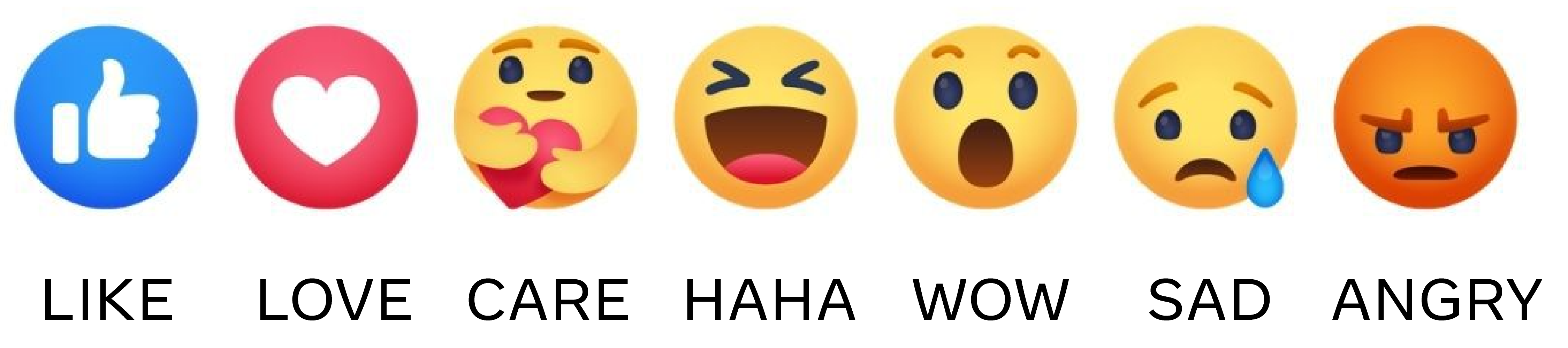}
  \caption{Facebook reactions.}
  \label{fig:reactions}
\end{figure}

\begin{figure}[H]
  \centering
  \includegraphics[width=0.6\linewidth]{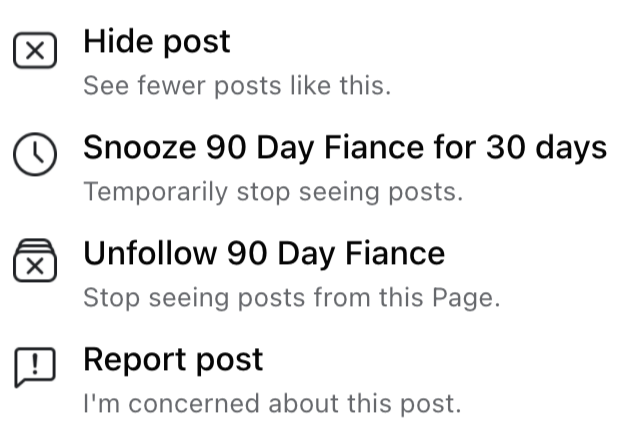}
  \caption{Facebook negative user feedback controls.}
  \label{fig:nuf}
\end{figure}

\begin{figure}[ht]
      \centering
  \includegraphics[width=0.75\linewidth]{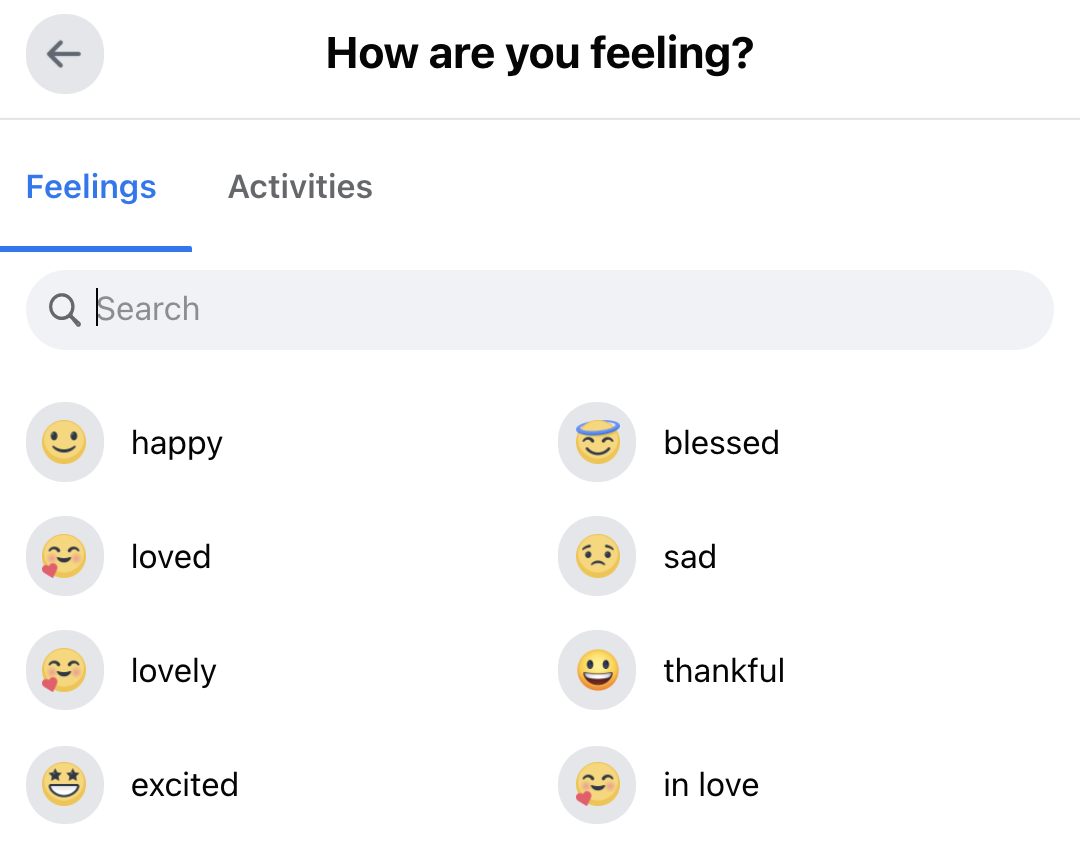}
  \caption{Facebook feature allowing posters to annotate their own posts with feelings.}
  \label{fig:minutiae}
\end{figure}

\section{Correlation between affective response and other Facebook signals}

\begin{figure*}
    \centering
  \includegraphics[width=0.63\linewidth]{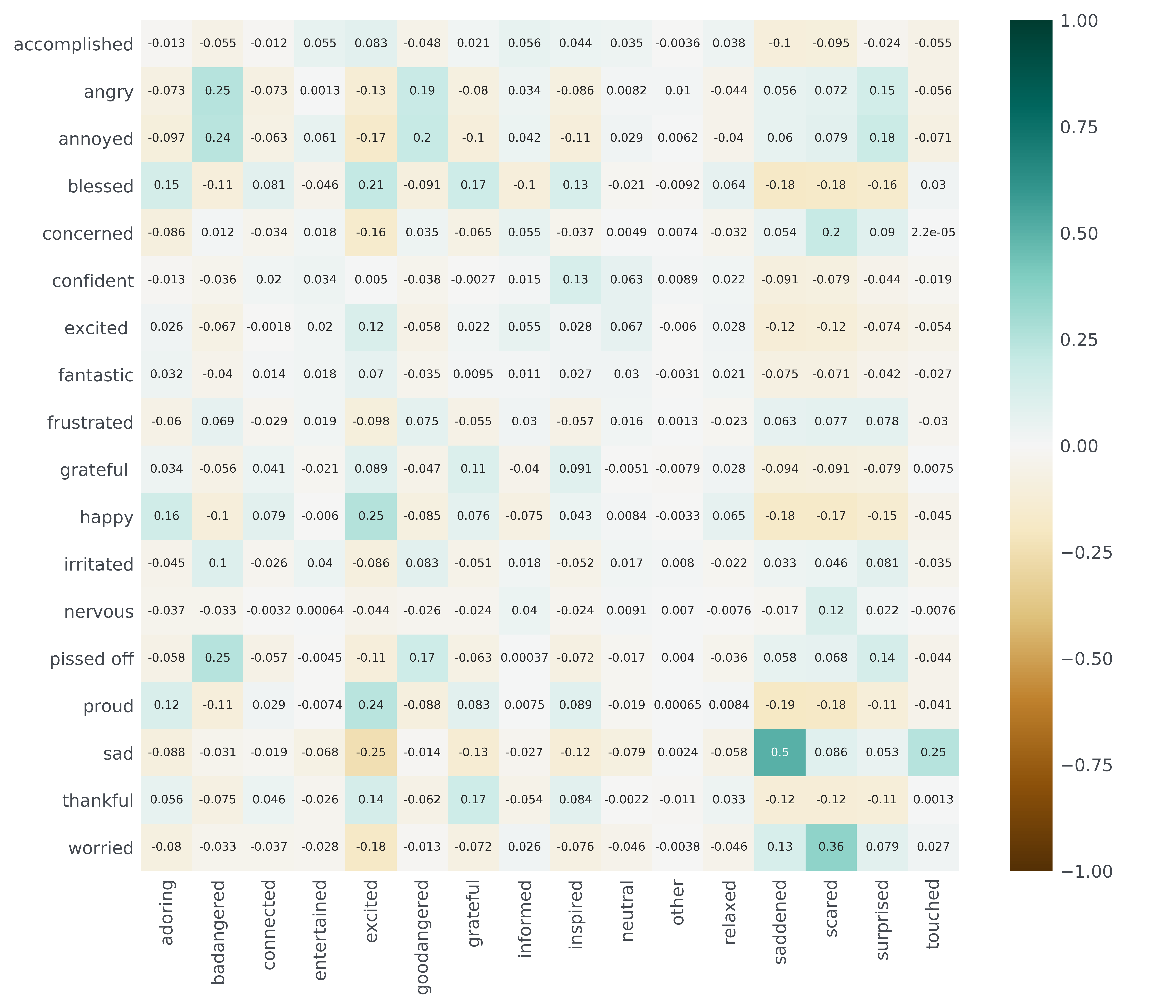} 
  \caption{Pearson correlation between human annotations for affective response and poster-annotated feelings.}
  \label{fig:minutiae_correlation}
\end{figure*}

\begin{figure*}
  \centering
   \includegraphics[width=0.63\linewidth]{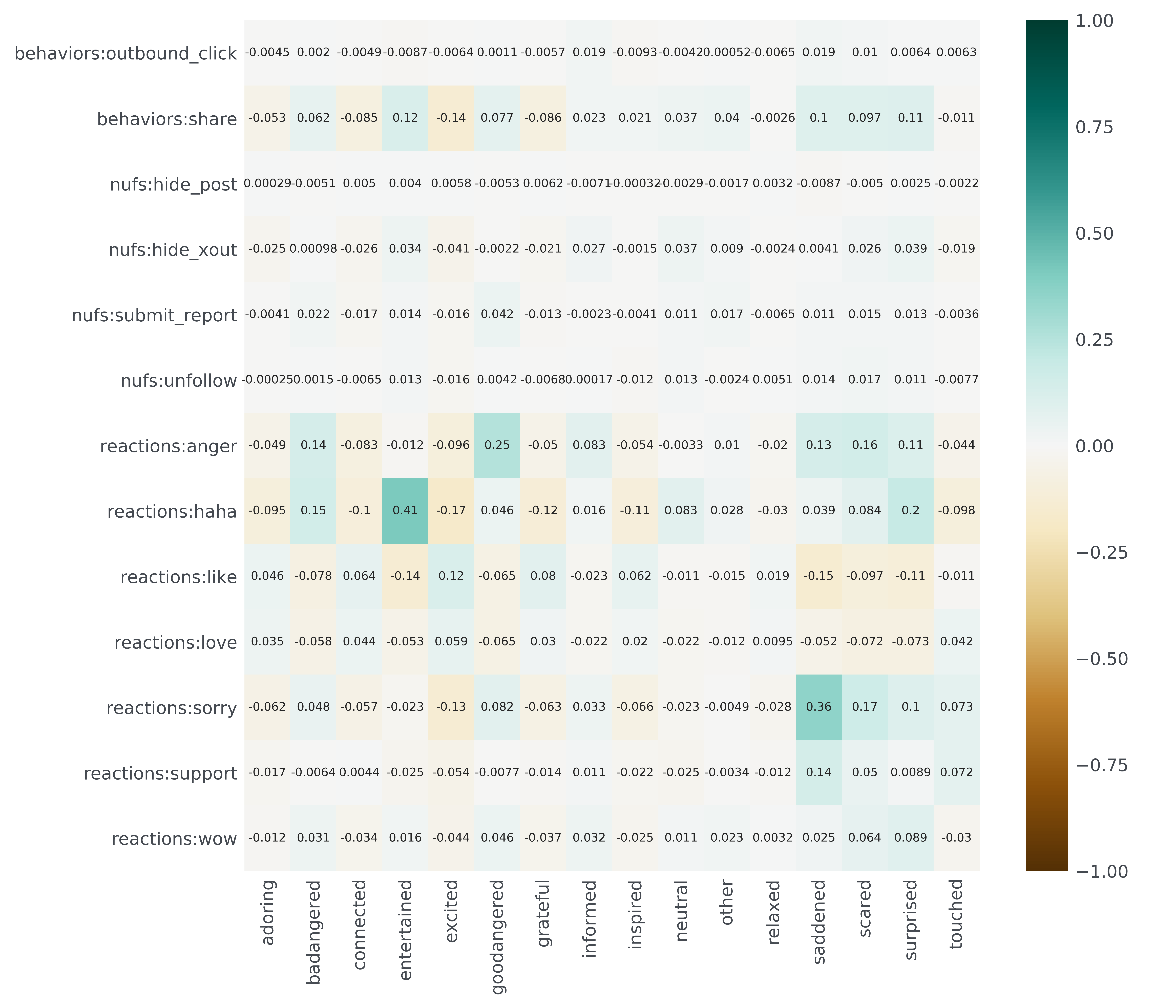} 
  \caption{Pearson correlation between human-annotations for affective response and user engagement signals. Values are first normalized by engagement type (i.e., behaviors, negative user feedback, and reactions).}
  \label{fig:affect_engagement}
\end{figure*}

In this section, we show the correlation between affective response and poster-annotated feelings in Figure~\ref{fig:minutiae_correlation} and between affective response and engagement signals in Figure~\ref{fig:affect_engagement}.


\end{document}